\documentclass[reprint,aps,prl,twocolumn,superscriptaddress,nofootinbib]{revtex4-2}

\usepackage{amsmath,amssymb,bm,graphicx}
\usepackage[dvipsnames]{xcolor}
\usepackage{times}
\usepackage[colorlinks=true,linkcolor=MidnightBlue,citecolor=MidnightBlue,urlcolor=MidnightBlue]{hyperref}

\begin{document}

\title{Impurity Self-Trapping in Lattice Bose systems}

\author{Chao Zhang}
\email{chaozhang@ahnu.edu.cn}
\affiliation{Department of Physics, Anhui Normal University, Wuhu, 241002, China}


\begin{abstract}
We map out the global phase diagram of a single mobile impurity in the two-dimensional
Bose--Hubbard model, spanning the bath evolution from a compressible superfluid (SF)
to an incompressible Mott insulator (MI) and the full range of impurity--bath coupling.
Using sign-problem-free worm-algorithm quantum Monte Carlo, we identify two distinct
self-trapping mechanisms that organize the entire diagram.
In the compressible SF, increasing impurity-bath coupling $|U_{\mathrm{ib}}|$ drives an interaction-driven
self-trapping crossover signaled by a collapse of the \emph{impurity} winding number:
a light, extended polaron evolves continuously into a heavy polaron and ultimately into
a self-trapped state---a repulsive \emph{saturated bubble} or an attractive \emph{bound cluster}---even
while the bath remains globally superfluid, demonstrating self-trapping without any bath phase transition.
By contrast, when the bath is tuned across the SF–MI transition at fixed $U_{\mathrm{ib}}$, localization is compressibility controlled. The vanishing bath compressibility quenches long-wavelength density redistribution and suppresses polaronic dressing, converting the SF polaron into a weakly dressed, nearly free defect upon entering the MI when $|U_{\mathrm{ib}}| \le 8.0$. Then increasing $|U_{\mathrm{ib}}|$ triggers a distinct Mott-specific route: the impurity binds a quantized vacancy or particle excitation, manifested by discrete changes $\Delta N_b=\pm1$ in the total bath occupation.
Together, our results provide a unified microscopic picture of impurity self-trapping in correlated lattice bosons,
governed by winding collapse in the SF and by compressibility loss and defect quantization across the SF--MI boundary.
\end{abstract}

\maketitle

\noindent\textbf{Introduction--}
A single mobile impurity embedded in a quantum many-body medium provides a minimal yet stringent
probe of correlations, coherence, and transport.
In weakly interacting Bose gases, the impurity forms a well-defined Bose polaron: a coherent
quasiparticle with finite residue, a renormalized effective mass, and an extended dressing cloud
built from gapless density fluctuations\cite{Grusdt_2025, ScazzaZaccanti2022, ArdilaGiorgini2015, ArdilaGiorgini2016, Ardila2019, Ardila2020, PhysRevLett.120.050405, PhysRevA.109.013325, PhysRevA.98.063631, 9ch1-dnvc}.
A central question is how this quasiparticle picture breaks down as correlations grow.
In particular, can an impurity lose mobility and self-trap \emph{purely through interactions} even in an
otherwise uniform, translation-invariant superfluid, and how does this fate change once the bath
itself becomes incompressible?

The two-dimensional Bose--Hubbard model at \emph{unit filling} ($\langle n_{\mathrm b}\rangle=1$) provides an ideal
setting to address these questions~\cite{JakschPRL1998, PhysRevB.40.546, PhysRevB.75.134302, CapogrossoSansonePRA2008}.
By tuning the intra-bath interaction $U_{\mathrm b}/t$, the bath evolves from a compressible superfluid (SF)
to an incompressible Mott insulator (MI).
Adding a single impurity with a local impurity--bath coupling $U_{\mathrm{ib}}$ (repulsive or attractive)
poses a sharply defined fate problem:
\emph{what are the distinct regimes of impurity motion and self-trapping across the full
$(U_{\mathrm b}/t,U_{\mathrm{ib}}/t)$ parameter space, and what microscopic mechanisms separate them?}

While the limiting regimes are qualitatively understood~\cite{10.21468/SciPostPhys.19.1.002, Santiago-García_2024, DuttaMueller2013, ColussiPRL2023, KeilerNJP2020, DingSciPost2023, Hartweg2025, PhysRevB.94.220502},
their global connection has remained incomplete.
In the compressible SF, most studies emphasize the coherent polaron regime, where the
impurity remains mobile and its dressing is supported by long-wavelength density fluctuations.
In the deep MI, the bath has a gap and vanishing compressibility; at weak coupling the impurity is
nearly a free defect with minimal dressing, whereas at strong coupling it can bind to and pin a
vacancy- or particle-type defect.
However, a unified microscopic account of \emph{how} impurity transport and dressing are lost---either by
increasing $|U_{\mathrm{ib}}|$ within the SF or by quenching bath compressibility across the SF--MI transition---is still lacking.
In particular, interaction-driven self-trapping in a uniform SF background \emph{without any bath phase transition}
has not been established in a controlled, unbiased setting.

Here we resolve this problem by constructing the \emph{global phase diagram} of a single mobile impurity
in the two-dimensional Bose--Hubbard model at $\langle n_{\mathrm b}\rangle=1$.
We employ large-scale, sign-problem-free worm-algorithm quantum Monte Carlo~\cite{ProkofevJETP1998, secondworm, PhysRevA.81.053622, Lingua_2018}
to obtain unbiased equilibrium results across both SF and MI regimes.
Crucially, we go beyond characterizing the impurity alone and track the \emph{microscopic bath response}---the impurity-induced
density redistribution and its collapse as the bath loses compressibility.
Our diagnostics quantify impurity transport via \emph{impurity winding statistics}, impurity coherence via single-particle correlations,
and the bath response via impurity-centered density modulations; in the MI we further identify \emph{quantized} vacancy/particle
defects through discrete changes in the total bath occupation.

The resulting phase diagram is organized by two sharply distinct self-trapping mechanisms.
\emph{Within the compressible SF}, increasing $|U_{\mathrm{ib}}|$ drives an
\emph{interaction-driven winding-collapse crossover}:
a light, extended polaron with finite impurity winding evolves continuously into a heavy polaron and ultimately
self-traps into a compact bound state---a repulsive \emph{saturated bubble} or an attractive \emph{bound cluster}---even
while the bath remains globally superfluid.
This establishes self-trapping without any bath phase transition, controlled by an impurity-winding collapse. \emph{Across the SF--MI transition at fixed $U_{\mathrm{ib}}$}, localization becomes \emph{compressibility controlled}:
as the bath compressibility is quenched, polaronic dressing and the impurity-induced density response collapse,
converting the SF polaron into a weakly dressed, nearly free defect on entering the MI when $|U_{\rm{ib}}| \le 8.0$.
Further increasing $|U_{\mathrm{ib}}|$ drives a distinct \emph{quantized-defect} regime,
where the impurity binds and pins a vacancy (repulsive) or particle (attractive) excitation, signaled by discrete
$\Delta N_b=\pm1$ changes in the total bath occupation.

Together, these results provide a unified microscopic framework for impurity self-trapping in correlated lattice bosons,
governed by winding collapse in the SF and by compressibility loss and defect quantization across the SF--MI boundary.

\noindent\textbf{Model and observables--}
We consider the two-dimensional Bose--Hubbard model with a single distinguishable mobile impurity coupled to a bosonic bath,
\begin{align}
H = & 
-t_{\mathrm{imp}} \!\!\sum_{\langle i,j\rangle}
   (a_i^\dagger a_j + \mathrm{H.c.})
 + U_{\mathrm{ib}}\sum_i n_{\mathrm{imp},i} n_{\mathrm{b},i} \nonumber \\
&  -t_{\mathrm{b}} \!\!\sum_{\langle i,j\rangle}
   (b_i^\dagger b_j + \mathrm{H.c.})
 + \frac{U_{\mathrm{b}}}{2}\sum_i n_{\mathrm{b},i}(n_{\mathrm{b},i}-1) \nonumber \\
& - \mu_{\mathrm{b}}\sum_i n_{\mathrm{b},i} ,
\label{eq:Hamiltonian}
\end{align}
Here $b_i$ ($a_i$) annihilates a bath (impurity) boson on site $i$,
$n_{{\mathrm b},i}=b_i^\dagger b_i$ and $n_{{\mathrm{imp}},i}=a_i^\dagger a_i$ are the corresponding number operators,
and $t_{\mathrm b}$ ($t_{\mathrm{imp}}$) denotes the nearest-neighbor hopping amplitude of the bath (impurity).
We restrict to the single-impurity limit $\sum_i n_{{\mathrm{imp}},i}=1$ and focus on unit filling of the bath,
$\langle n_{{\mathrm b},i}\rangle=1$. In the superfluid regime, the bath chemical potential $\mu$ is tuned (grand-canonical ensemble) to reach
integer filling 1 within numerical accuracy.
In the Mott-insulating regime, we fix the chemical potential to the strong-coupling estimate for the center of the lobe of integer filling 1 on the square lattice, $\mu_{\rm b} = (\sqrt{2}-1)\,U_{\mathrm b}$, so that the bath is placed close to the gap midpoint.

The on-site interaction $U_{\mathrm{b}}$ tunes the bath from a compressible superfluid (SF) to an incompressible Mott insulator (MI),  
while the impurity--bath coupling $U_{\mathrm{ib}}$ can be either repulsive ($U_{\mathrm{ib}}>0$) or attractive ($U_{\mathrm{ib}}<0$).

We set $t_{\mathrm{b}}=t_{\mathrm{imp}}=t=1$ as the energy unit and vary $U_{\mathrm{b}}/t$ from $8.0$ (deep SF) to $24.0$ (deep MI).  All simulations are performed on $L\times L$ lattices (typically $L=20$) at inverse temperature $\beta=L$ with periodic boundary condition,  
which ensures convergence to the ground-state regime.

\begin{figure}[t]
    \centering
    \includegraphics[width=\linewidth]{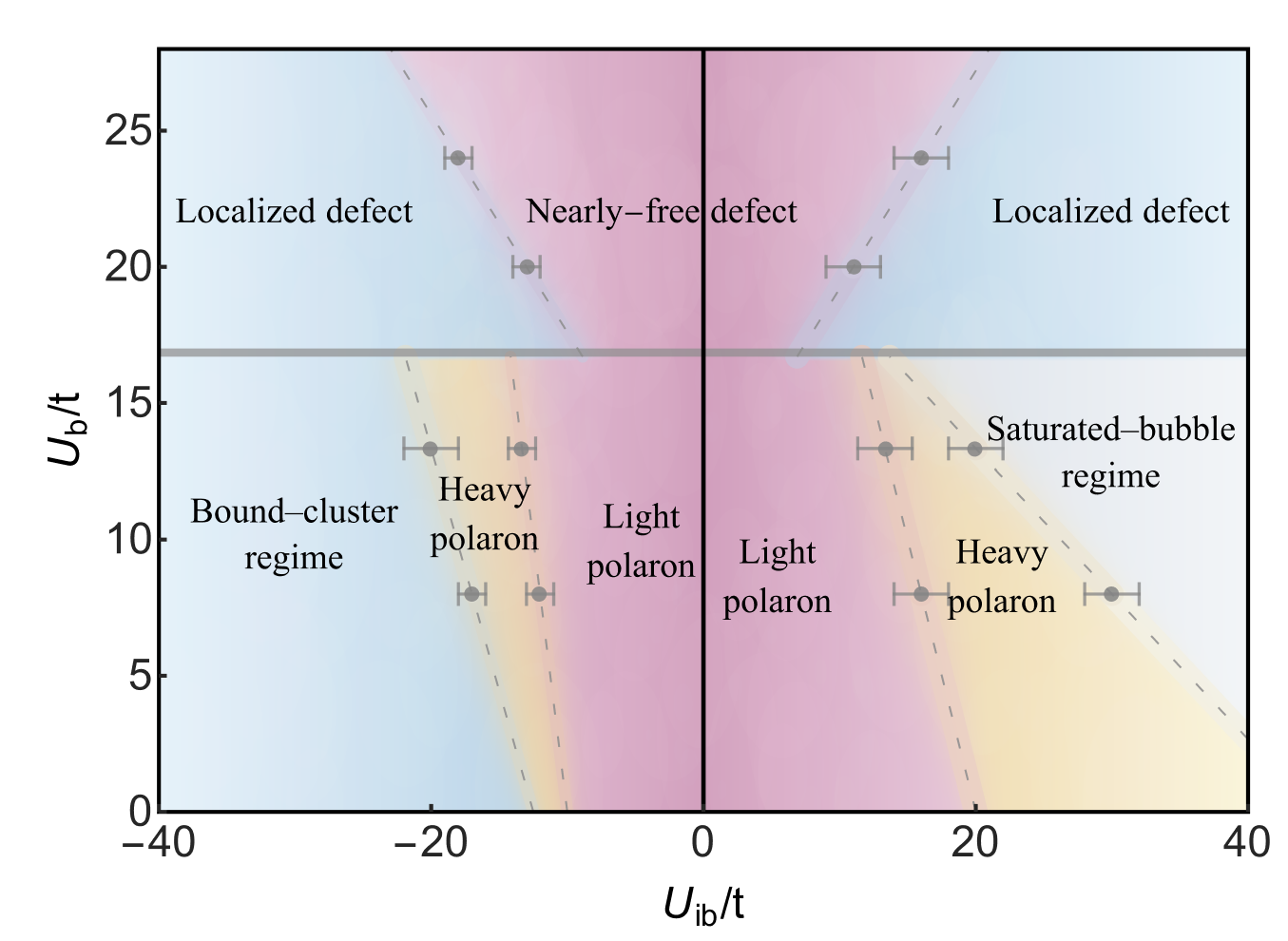}
\caption{
\textbf{Phase diagram of a single impurity in the two-dimensional Bose--Hubbard model
in the $(U_{\mathrm{ib}}/t,\, U_{\mathrm{b}}/t)$ plane.}
The vertical dashed line separates repulsive ($U_{\mathrm{ib}}/t>0$) and attractive
($U_{\mathrm{ib}}/t<0$) impurity--bath couplings, while the horizontal dashed line marks
the superfluid--Mott transition of the bath ($U_{\mathrm{b}}/t\simeq16.7$).
In the \emph{compressible superfluid} region ($U_{\mathrm{b}}/t<16.7$), increasing
$|U_{\mathrm{ib}}|/t$ drives an \emph{interaction-driven self-trapping crossover} signaled by a collapse of the
\emph{impurity} winding number: a mobile light polaron evolves continuously into a heavy polaron and,
at strong coupling, into a saturated bubble (repulsive side) or a bound cluster (attractive side).
In the \emph{incompressible Mott-insulating} regime ($U_{\mathrm{b}}/t\gtrsim16.7$),
self-trapping proceeds via \emph{quantized defect formation}.
Beyond a threshold in $|U_{\mathrm{ib}}|/t$, the impurity changes abruptly from a nearly free defect
to a pinned vacancy (repulsive) or particle (attractive) defect, signaled by an integer jump
$\Delta N_{\mathrm b}=\mp1$ in the total bath occupation.
For intermediate couplings, the impurity remains weakly dressed and essentially free.
Along vertical cuts at fixed $U_{\mathrm{ib}}/t$, crossing the SF--MI boundary thus realizes
\emph{compressibility-controlled self-trapping}.
Grey dots mark QMC data points; solid lines are guides to the eye
highlighting continuous crossover boundaries in the superfluid, while dotted
lines indicate sharp transitions between distinct defect regimes in the Mott
phase.}
    \label{fig:phase_diagram}
\end{figure}

\emph{Polaron properties--}
We access impurity quasiparticle properties via the single-impurity Green's function measured within the worm algorithm. We accumulate the real-space impurity Green's function $G_{\mathrm{imp}}(\mathbf r,\tau)=\langle a(\mathbf r,\tau)\,a^{\dagger}(\mathbf 0,0)\rangle$ and obtain the momentum-resolved Green's function by a discrete Fourier transform, $G_{\mathrm{imp}}(\mathbf k,\tau)=\sum_{\mathbf r} e^{-i\mathbf k\cdot\mathbf r}\,G_{\mathrm{imp}}(\mathbf r,\tau)$. At low temperature and large imaginary time, $G_{\mathrm{imp}}(\mathbf k,\tau)$ is dominated by the lowest impurity state, $G_{\mathrm{imp}}(\mathbf k,\tau)\simeq Z_{\mathbf k} e^{-E_{\mathrm{p}}(\mathbf k)\tau}$, from which we extract the polaron energy $E_{\mathrm{p}}(\mathbf k)$ and quasiparticle residue $Z_{\mathbf k}$. The effective mass $m^*/m_0=2t/\frac{\partial^2E_{\mathrm{p}}(\mathbf{k})}{\partial \mathbf{k}^2}$ is obtained from the small-momentum dispersion by fitting the lowest available momenta. 

\emph{Winding-number diagnostics--}
To characterize impurity mobility and the global coherence of the bath, we measure the winding numbers of both components.
For the impurity, we record the winding number square of the impurity
$\langle W_{\mathrm {imp}}^2\rangle=\langle W_{\mathrm{imp},x}^2+W_{\mathrm{imp},y}^2\rangle$,
where $W_{\mathrm{imp}, x}$ and $W_{\mathrm{imp},y}$ denote the net worldline crossings through the $x$ and $y$ directions.
A finite $\langle W_{\mathrm {imp}}^2\rangle$ signals a mobile impurity with extended trajectories, whereas a strong suppression of $\langle W_{\mathrm{imp}}^{2}\rangle$ toward zero signals an effectively self-trapped impurity. 
For the bath, the corresponding estimator gives the superfluid density
$\rho_{\mathrm {b}}=\langle W_{{\mathrm {b}},x}^2+W_{{\mathrm {b}},y}^2\rangle/(2t_{\rm b}\beta)$,
directly related to the superfluid density via the Pollock--Ceperley relation~\cite{PhysRevB.36.8343}. 

\emph{Bath response diagnostics--}
To quantify the spatial response of the bath to a single impurity, we evaluate the impurity-centered correlator $C_{\mathrm{ib}}(\mathbf r)=\sum_{\mathbf r_{\mathrm{imp}}} w(\mathbf r_{\mathrm{imp}})\,[\langle n_{\mathrm{b}}(\mathbf r_{\mathrm{imp}}+\mathbf r)\rangle-\bar n_{\mathrm{b}}]$,
which measures the bath density change at displacement $\mathbf r$ from the instantaneous impurity position $\mathbf r_{\mathrm{imp}}$.
Here $\bar n_{\mathrm{b}}=L^{-2}\sum_i\langle n_{\mathrm{b},i}\rangle$ is the uniform bath density.  Because the impurity is a quantum object whose worldline freely explores the lattice in imaginary time, its position is not fixed.  
The weighting function
$w(\mathbf r_{\mathrm{imp}})=\beta^{-1}\!\int_0^\beta\!\langle n_{\mathrm {imp}}(\mathbf r_{\mathrm{imp}},\tau)\rangle\,d\tau$
represents the time-averaged probability of finding the impurity at site $\mathbf r_{\mathrm{imp}}$, satisfying $\sum_{\mathbf r_{\mathrm{imp}}}w(\mathbf r_{\mathrm{imp}})=1$.

To remove lattice anisotropy, we perform a rotational averaging over all lattice sites with the same Euclidean distance
$R=|\mathbf r|=\sqrt{r_x^2+r_y^2}$ (using the minimum–image convention), defining
$C_{\mathrm{ib}}(R)=N_R^{-1}\!\sum_{|\mathbf r|=R} C_{\mathrm{ib}}(\mathbf r)$,
where $N_R$ counts sites satisfying $|\mathbf r|=R$.
This smooth radial profile characterizes the impurity-centered bath density deformation:
$C_{\mathrm{ib}}(R)<0$ signals a local depletion (\emph{bubble}) for repulsive $U_{\mathrm{ib}}/t>0$,
while $C_{\mathrm{ib}}(R)>0$ indicates a density accumulation (\emph{cluster}) for attractive $U_{\mathrm{ib}}/t<0$.
The cumulative excess
$\Delta N(R)=\sum_{|\mathbf r|\le R}C_{\mathrm{ib}}(\mathbf r)$
gives the net bath particle-number change within a disk of radius $R$ centered at the impurity, whose amplitude and extent distinguish the light, heavy, and self-trapped regimes discussed below.

\begin{figure}[t]
    \centering
    \includegraphics[width=\linewidth]{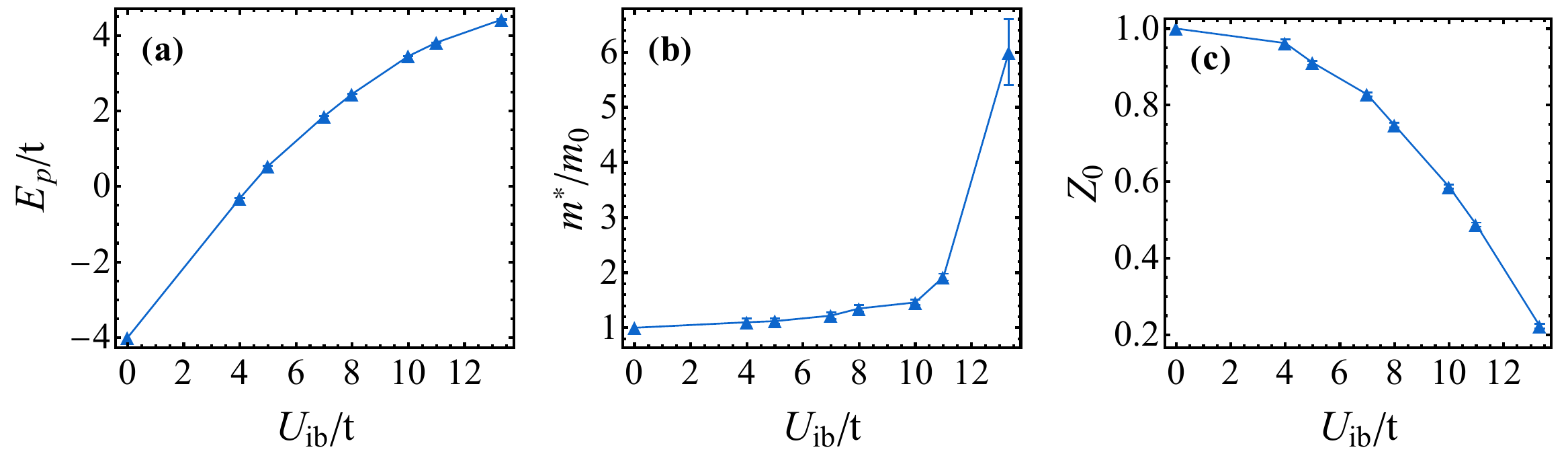}
\caption{
\textbf{Impurity quasiparticle properties in a superfluid bath
($U_{\mathrm{b}}/t = 13.3$).}
(a)~Ground-state energy $E_{\mathrm{p}}/t$, (b)~effective-mass ratio $m^*/m_0$, and
(c)~quasiparticle residue $Z_0$ versus the repulsive impurity--bath coupling $U_{\mathrm{ib}}/t$.
Here $m_0$ is the bare lattice mass of the impurity (nearest-neighbor hopping).
The quasiparticle diagnostics $m^*$ and $Z_0$ are shown in the mobile regime with finite impurity winding,
$\langle W_{\mathrm{imp}}^2\rangle>0$.
Increasing $U_{\mathrm{ib}}$ continuously enhances mass renormalization and suppresses $Z_0$,
signaling stronger impurity--bath correlations.
At $U_{\mathrm{ib}}/t \gtrsim 13.3$, the impurity winding collapses to zero
[Fig.~\ref{fig:realspace}(e)], and the quasiparticle description breaks down, marking the boundary of the
light-polaron regime.
The smooth evolution of $E_{\mathrm{p}}/t$ indicates a continuous interaction-driven crossover rather than a sharp transition.}
    \label{fig:polaron_properties}
\end{figure}

Together, these observations reveal a clear dichotomy in how a single impurity self-traps in the
two-dimensional Bose--Hubbard model, as summarized in Fig.~\ref{fig:phase_diagram}.
In the compressible superfluid bath ($U_{\mathrm{b}}/t \lesssim 16.7$), self-trapping is continuous:
increasing $|U_{\mathrm{ib}}|/t$ progressively suppresses coherent impurity winding and compresses the dressing cloud,
yielding a smooth crossover from light to heavy polarons and, at strong coupling, to a saturated bubble (repulsive)
or a bound cluster (attractive) state.
We refer to this route as \emph{interaction-driven, winding-collapse self-trapping}.

In contrast, along vertical cuts at fixed $U_{\mathrm{ib}}/t=8.0$, increasing $U_{\mathrm{b}}/t$ across the SF--MI transition
renders the impurity evolution \emph{compressibility controlled}:
as the bath compressibility is quenched, long-wavelength density redistribution and polaronic dressing collapse, and the SF polaron
is converted into a weakly dressed, nearly free defect on entering the MI.
\emph{Within the incompressible MI}, further increasing $|U_{\mathrm{ib}}|/t$ drives a distinct quantized-defect regime in which the impurity
binds and pins a vacancy (repulsive) or particle (attractive) excitation, signaled by an integer bath-occupation change
$\Delta N_{\mathrm b}=\pm1$.
We refer to this two-stage mechanism as \emph{compressibility-controlled self-trapping}.

These two routes—a continuous, interaction-driven winding-collapse mechanism in the superfluid and a compressibility-controlled route that converts the impurity into a weakly dressed, nearly free defect on entering the Mott insulator at intermediate coupling—define two fundamentally distinct paradigms of impurity self-trapping in strongly correlated lattice bosons.

\begin{figure}[t]
    \centering
\includegraphics[width=\linewidth, trim = 0pt 0pt 430pt 0pt,  clip]{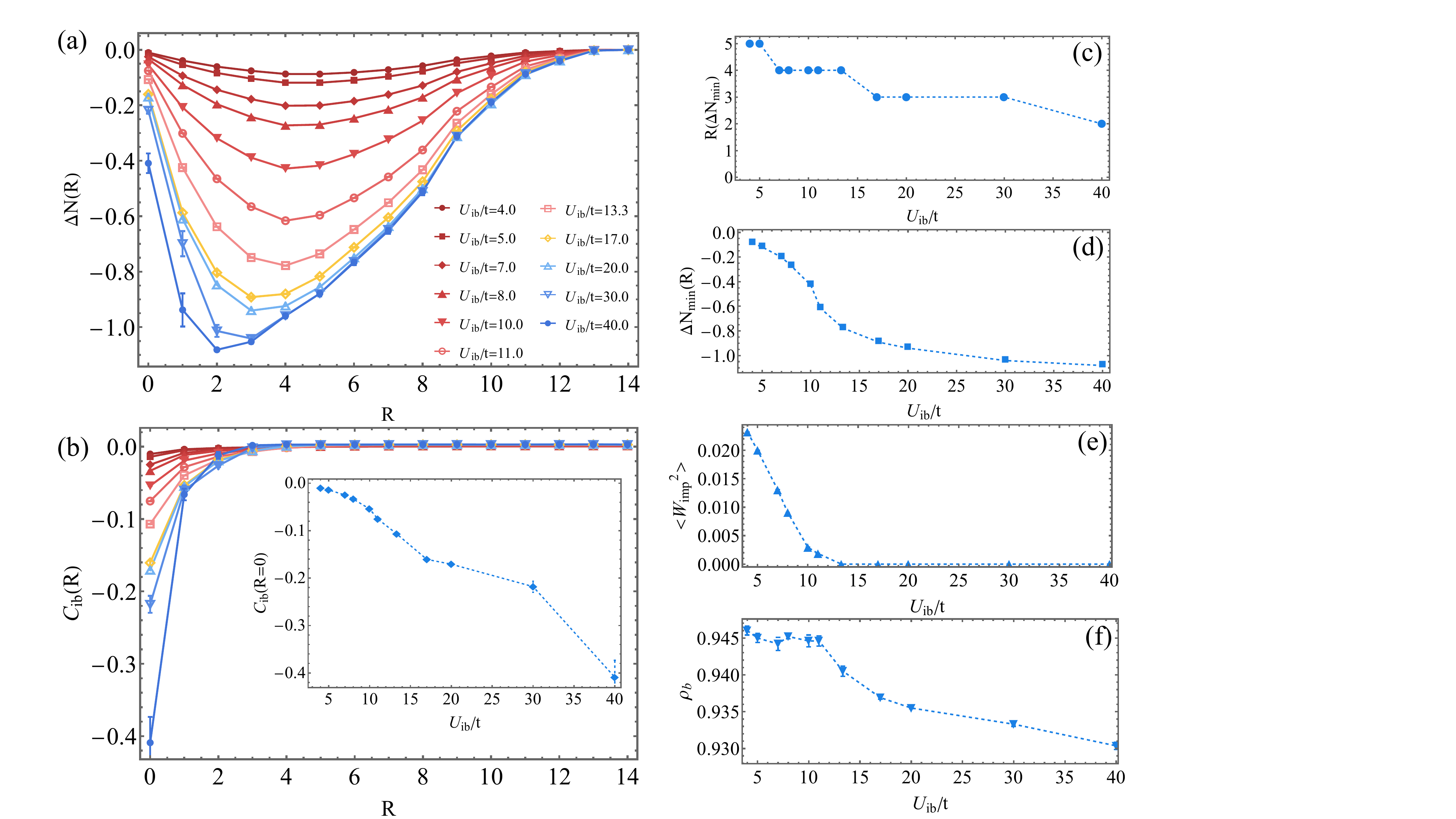}
\caption{
\textbf{Real-space crossover from a light polaron to a saturated bubble in a superfluid bath ($U_{\mathrm{b}}/t=13.3$).}
(a)~Cumulative bath-density response $\Delta N(R)$ for increasing repulsive impurity--bath coupling $U_{\mathrm{ib}}/t$.
Red curves denote the light-polaron regime, the orange curve highlights an intermediate heavy-polaron regime,
and blue curves correspond to the saturated-bubble regime where the impurity is self-trapped.
(b)~Impurity-centered correlator $C_{\mathrm{ib}}(R)$ and its on-site value $C_{\mathrm{ib}}(0)$ (inset).
(c,d)~Position $R_{\min}$ and depth $\Delta N_{\min}$ of the depletion minimum extracted from $\Delta N(R)$.
(e)~Impurity winding $\langle W_{\mathrm{imp}}^2\rangle$, showing a continuous loss of coherent motion.
(f)~Bath superfluid density $\rho_{\mathrm{b}}$, which remains nearly unchanged across the crossover.
The color progression illustrates a continuous evolution from a mobile light polaron (red) to an immobile but still dressed heavy polaron (orange),
and finally to a fully self-trapped saturated bubble (blue).}
    \label{fig:realspace}
\end{figure}

\noindent\textbf{Interaction-driven polaron self-trapping in a superfluid bath:}
\textit{(i) Quasiparticle renormalization in the light-polaron regime.}
To characterize the impurity in a compressible superfluid background,
we fix the bath interaction at $U_{\mathrm{b}}/t = 13.3$ (deep in the SF phase)
and increase the repulsive impurity--bath coupling $U_{\mathrm{ib}}/t>0$.
Figure~\ref{fig:polaron_properties} shows the ground-state energy $E_{\mathrm{p}}/t$,
the effective-mass ratio $m^*/m_0$, and the quasiparticle residue $Z_0$ extracted from
imaginary-time Green's functions.
The quasiparticle quantities are reported in the mobile regime with finite impurity winding
$\langle W_{\mathrm{imp}}^2\rangle>0$, where a coherent polaron is well defined.

At weak coupling ($U_{\mathrm{ib}}/t \lesssim 10.0$), the impurity forms a light polaron:
$m^*/m_0 \lesssim 2.0$ and $Z_0\gtrsim 0.5$, with $E_{\mathrm{p}}/t$ varying smoothly,
consistent with relatively weak dressing.
For intermediate coupling ($10.0\lesssim U_{\mathrm{ib}}/t\lesssim13.3$),
$m^*/m_0$ increases and $Z_0$ decreases continuously, reflecting enhanced impurity--bath correlations
while coherent transport persists.
Beyond $U_{\mathrm{ib}}/t\gtrsim13.3$, the impurity winding collapses to zero
[Fig.~\ref{fig:realspace}(e)], signaling the loss of coherent motion and the breakdown of the quasiparticle picture:
$m^*$ rapidly grows (effectively diverges as $\langle W_{\mathrm{imp}}^2\rangle\!\to\!0$) and $Z_0$ drops toward zero.
Throughout the mobile regime, $E_{\mathrm{p}}/t$ evolves smoothly with $U_{\mathrm{ib}}/t$,
confirming that the dressing and eventual self-trapping develop via a continuous crossover rather than a sharp transition.

\textit{(ii) Real-space crossover from light polaron to heavy polaron and saturated bubble.}
To elucidate how the impurity reshapes its surrounding superfluid environment,
we fix the bath interaction at $U_{\mathrm{b}}/t=13.3$ (deep in the SF regime) and increase the repulsive impurity--bath coupling $U_{\mathrm{ib}}/t>0$.
Figure~\ref{fig:realspace} summarizes the corresponding observables and reveals a continuous evolution from a mobile \emph{light polaron} (red curves),
to an \emph{heavy polaron} (orange), and finally to a spatially \emph{saturated bubble} (blue).

Figure~\ref{fig:realspace}(a) shows the cumulative bath-density response $\Delta N(R)$.
At weak coupling ($U_{\mathrm{ib}}/t\lesssim 13.3$), $\Delta N(R)$ exhibits a shallow, broad minimum around
$R_{\min}\approx 4$--5 with $\Delta N_{\min}\gtrsim -0.4$, consistent with an extended depletion cloud accompanying a mobile light polaron.
Upon increasing $U_{\mathrm{ib}}/t$ into the intermediate range ($13.3\lesssim U_{\mathrm{ib}}/t\lesssim 20$),
the deformation deepens and contracts to $R_{\min}\approx 2$--3 while $\Delta N_{\min}$ approaches $-0.8$.
At the same time, the impurity winding $\langle W_{\mathrm{imp}}^2\rangle$ drops to (nearly) zero
[Fig.~\ref{fig:realspace}(e)], indicating that coherent transport is lost even though a finite-range dressing cloud persists.
We identify this as a \emph{heavy-polaron} regime: the impurity is effectively immobile, yet still surrounded by a compact many-body deformation.

For stronger coupling ($U_{\mathrm{ib}}/t\gtrsim 20$), the $\Delta N(R)$ curves become nearly coupling independent for $R\gtrsim 3$--4,
approaching a common spatial profile whose extent no longer shrinks appreciably with $U_{\mathrm{ib}}/t$.
This marks the onset of a \emph{saturated-bubble} regime: the depletion cloud has reached a saturated \emph{spatial} size, while its depth continues to evolve with $U_{\mathrm{ib}}/t$.
Importantly, this saturation is not quantized (unlike the integer defects in the incompressible Mott background; see~\cite{chaoPRB}).

Figure~\ref{fig:realspace}(b) shows the impurity-centered correlator $C_{\mathrm{ib}}(R)$, whose suppression at short distances (inset) tracks the same crossover.
For $U_{\mathrm{ib}}/t\gtrsim 20$, $C_{\mathrm{ib}}(R\!>\!1)$ becomes weakly dependent on $U_{\mathrm{ib}}/t$, corroborating that the dressing cloud has reached a saturated spatial extent.
Panels (c,d) quantify the contraction and deepening of the deformation cloud through $R_{\min}$ and $\Delta N_{\min}$.
Meanwhile, $\langle W_{\mathrm{imp}}^2\rangle$ decreases smoothly by nearly two orders of magnitude across the light to heavy polaron regime
[Fig.~\ref{fig:realspace}(e)], consistent with a continuous crossover rather than a sharp phase transition.
In contrast, the bath superfluid density $\rho_{\mathrm{b}}$ remains finite and is only weakly affected
[Fig.~\ref{fig:realspace}(f)], confirming that the impurity self-traps through interaction-driven winding collapse while the bath stays globally superfluid.

Collectively, these observables delineate three regimes controlled solely by $U_{\mathrm{ib}}/t$:
(i)~a \emph{light-polaron} regime (red) with finite winding and a broad, weak depletion cloud;
(ii)~a \emph{heavy-polaron} regime (orange) where winding has collapsed but a compact finite-range deformation persists;
and (iii)~a \emph{saturated-bubble} regime (blue) where the deformation profile is spatially saturated.
This progression from mobile light polarons to a saturated bubble constitutes a continuous, non-singular self-trapping route driven purely by impurity--bath couplings within a compressible superfluid bath.

\noindent\textbf{Compressibility-controlled collapse of polaronic dressing}

\begin{figure}[t]
\centering
\includegraphics[width=\linewidth]{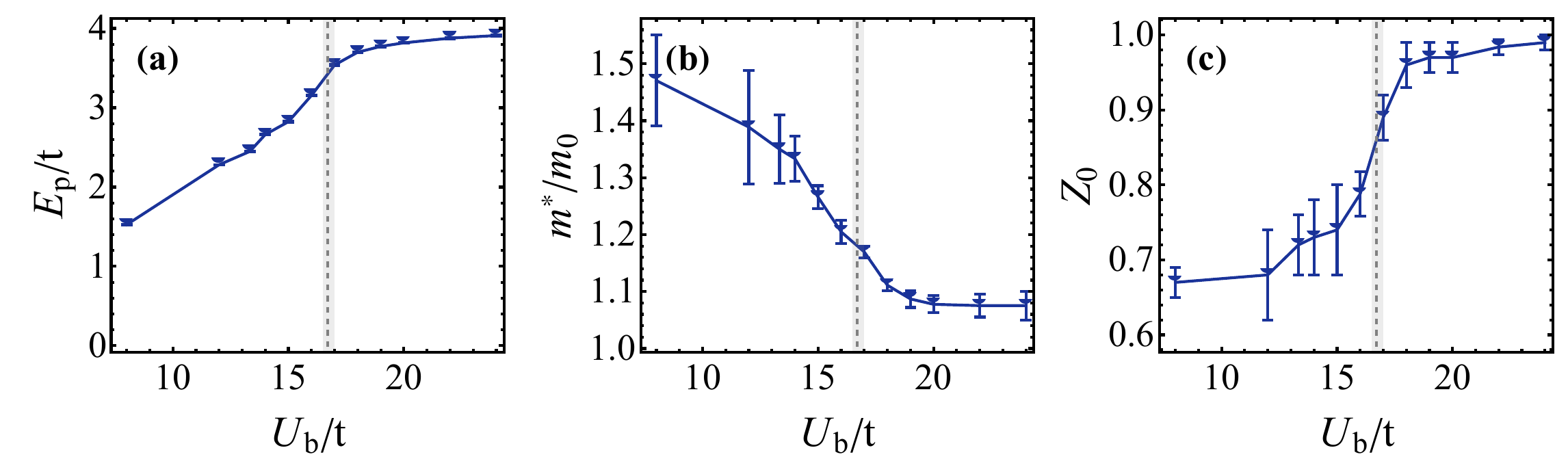}
\caption{
\textbf{Compressibility-controlled collapse of polaronic dressing at fixed repulsive impurity--bath coupling $U_{\mathrm{ib}}/t=8.0$.}
(a)~Ground-state energy $E_{\mathrm{p}}/t$, (b)~effective-mass ratio $m^*/m_0$, and (c)~quasiparticle residue $Z_0$
as functions of bath interaction $U_{\mathrm{b}}/t$.
The shaded region marks the bath SF--MI transition ($U_{\mathrm{b}}/t\simeq16.7$).
In the compressible SF ($U_{\mathrm{b}}/t<16.7$), the impurity forms a weakly renormalized light polaron with a finite dressing cloud.
As the bath becomes less compressible, $m^*/m_0$ decreases while $Z_0$ increases, indicating a progressive collapse of polaronic dressing.
Upon entering the incompressible MI ($U_{\mathrm{b}}/t\gtrsim16.7$), $m^*/m_0\to1$ and $Z_0\to1$,
signaling a nearly bare (almost free) impurity.}        \label{fig:compressibility_polaron}
        \end{figure}

\begin{figure}[t]
    \centering
\includegraphics[width=\linewidth, trim = 0pt 0pt 390pt 0pt,  clip]{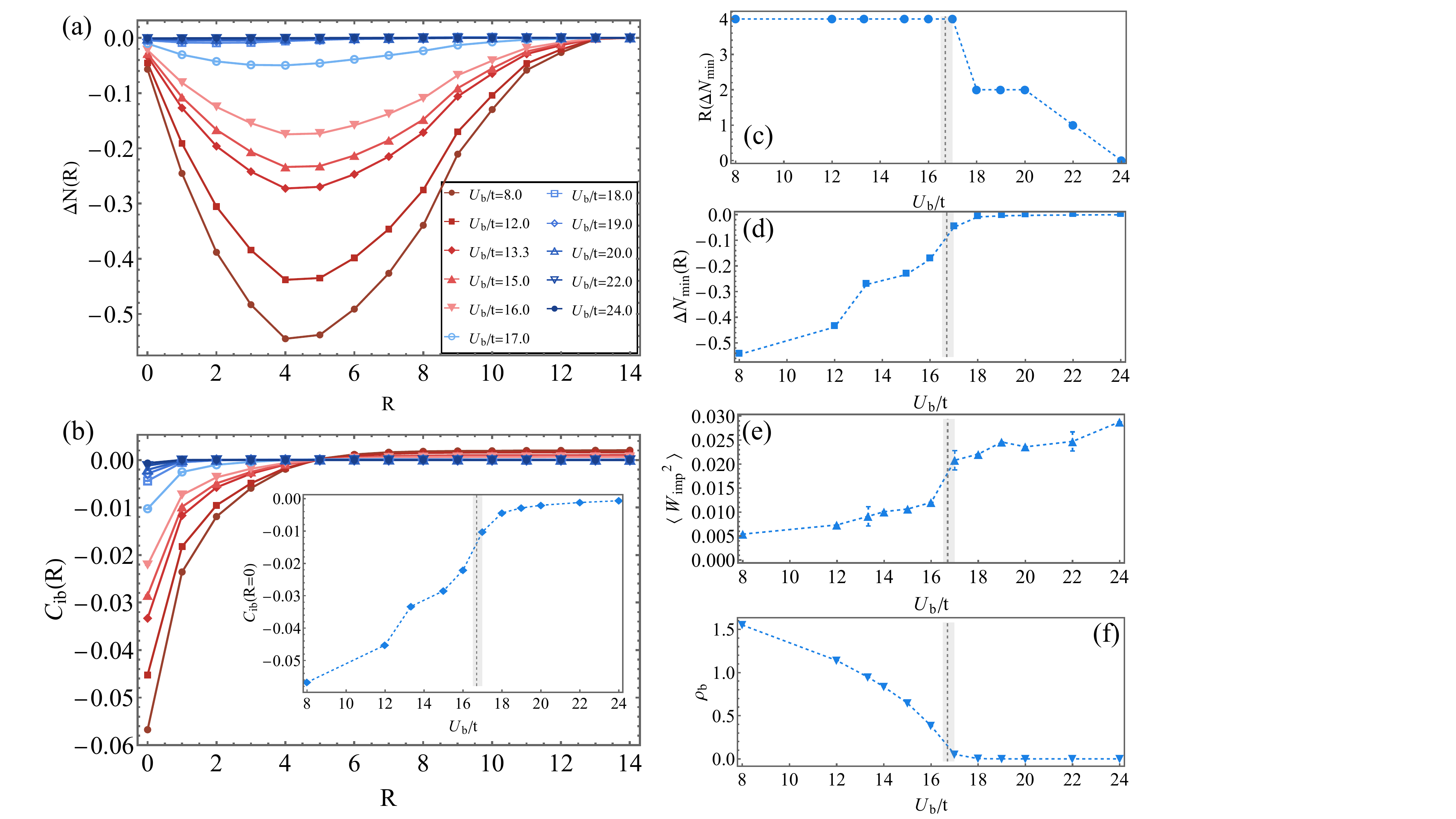}
\caption{
\textbf{Compressibility-controlled collapse of real-space dressing and defect binding at fixed $U_{\mathrm{ib}}/t=8.0$.}
(a)~Cumulative bath-density depletion $\Delta N(R)$ and (b)~impurity-centered correlator $C_{\mathrm{ib}}(R)$
for increasing $U_{\mathrm{b}}/t$, color-coded from deep SF (dark red: $U_{\mathrm{b}}/t=8.0$)
through intermediate (light red/orange: $U_{\mathrm{b}}/t=13.3,15.0$) to near the transition (light blue: $U_{\mathrm{b}}/t=17.0$)
and deep MI (dark blue: $U_{\mathrm{b}}/t=20.0,22.0,24.0$).
(c,d)~Minimum position $R_{\min}$ and depth $\Delta N_{\min}$ extracted from $\Delta N(R)$, quantifying the collapse of extended bath response.
(e)~Impurity winding $\langle W_{\mathrm{imp}}^{2}\rangle$ increases across the SF--MI transition, indicating enhanced coherent impurity transport.
(f)~Bath superfluid density $\rho_{\mathrm{b}}$ vanishes at the transition.
Together these observables reveal a compressibility-controlled crossover from a mobile polaron in the SF to an almost bare impurity upon entering the MI phase.}
    \label{fig:compressibility_collapse}
\end{figure}

\noindent\textit{(i) Polaron renormalization under decreasing bath compressibility.}
To isolate the role of the bath, we fix the impurity--bath coupling at a moderate repulsive value ($U_{\mathrm{ib}}/t=8.0$)
and vary $U_{\mathrm{b}}/t$ across the SF--MI transition.
Figures~\ref{fig:compressibility_polaron}(a--c) show that in the compressible SF ($U_{\mathrm{b}}/t<16.7$)
the impurity remains a well-defined polaron with finite winding
[Fig.~\ref{fig:compressibility_collapse}(e)].
As the bath becomes less compressible, the dressing weakens continuously: the effective-mass ratio $m^*/m_0$ decreases
(from $\sim1.5$ toward $\sim1.2$), while the residue $Z_0$ increases (from $\sim0.7$ toward $\sim0.9$),
indicating a progressively \emph{lighter} and more weakly dressed polaron.
Near the SF--MI boundary ($U_{\mathrm{b}}/t\simeq16.7$), $E_{\mathrm{p}}/t$ exhibits a visible change in slope and
the renormalization collapses, with $m^*/m_0\to1$ and $Z_0\to1$, signaling the recovery of an almost bare impurity.
Inside the incompressible MI, extended density redistribution is suppressed and the bath can no longer sustain polaronic correlations,
so the impurity behaves as a nearly free defect in the lattice background.

This compressibility-controlled \emph{depolaronization} is distinct from the interaction-driven self-trapping at large $|U_{\mathrm{ib}}|/t$ in the SF:
here the dressing collapses because the bath loses compressibility, not because the impurity--bath coupling is strong.

\noindent\textit{(ii) Real-space collapse and defect binding.}
Figures~\ref{fig:compressibility_collapse}(a,b) show how the real-space bath response evolves under the same protocol.
In the SF, $\Delta N(R)$ and $C_{\mathrm{ib}}(R)$ display an extended deformation cloud characteristic of a coherent polaron.
As $U_{\mathrm{b}}/t$ increases and the bath becomes less compressible, both the depth $\Delta N_{\min}$ and the extent $R_{\min}$ shrink continuously
[Figs.~\ref{fig:compressibility_collapse}(c,d)], reflecting the collapse of long-wavelength density response.
Concomitantly, the impurity winding $\langle W_{\mathrm{imp}}^2\rangle$ increases across the transition
[Fig.~\ref{fig:compressibility_collapse}(e)], while the bath superfluid density vanishes
[Fig.~\ref{fig:compressibility_collapse}(f)].
In the MI, the impurity is therefore only weakly dressed at intermediate $|U_{\mathrm{ib}}|/t$.
Upon further increasing $|U_{\mathrm{ib}}|/t$ within the incompressible regime, it binds and pins a vacancy-type defect for repulsive $U_{\mathrm{ib}}/t$,
signaled by a quantized depletion plateau in the integrated density response $\Delta N(R)$ approaching $-1$ (within our resolution).

This realizes the \emph{compressibility-controlled} route: polaronic dressing collapses as the bath becomes incompressible, and MI self-trapping proceeds via quantized defect binding.

\noindent\textbf{MI extension and outlook--}
In an \emph{incompressible MI} bath, the gapped spectrum and vanishing compressibility strongly suppress
long-wavelength density rearrangements, undermining mobile polarons and disfavoring extended bound states.
In the MI regime, at small to intermediate $|U_{\mathrm{ib}}|/t$ the impurity induces only short-range distortions, and quasiparticle diagnostics
($m^*/m_0$, $Z_0$, and $\langle W_{\mathrm{imp}}^{2}\rangle$) revert toward their bare-defect limits.
At sufficiently strong impurity--bath coupling, the impurity nucleates \emph{quantized} vacancy- or particle-type defects,
manifested by discrete changes $\Delta N_{\mathrm b}=\pm1$ in the total bath occupation.
The onset of this quantized-defect formation depends on the bath chemical potential $\mu_{\mathrm b}$ (i.e., the position within the Mott lobe),
which shifts the corresponding threshold lines without altering the underlying mechanism.
A systematic analysis of this insulating branch is reported in Ref.~\cite{chaoPRB}.

Several extensions are natural. Finite temperature will activate particle--hole excitations in the Mott background and can induce
a temperature-dependent \emph{delocalization} (re-mobilization) crossover of the impurity--defect complex.
Relaxing $t_{\mathrm {imp}}\!=\!t_{\mathrm b}$ adds a powerful knob: $t_{\mathrm {imp}}\!\ll\!t_{\mathrm b}$ approaches the static-defect limit,
whereas $t_{\mathrm {imp}}\!\gg\!t_{\mathrm b}$ realizes a fast impurity in a slow, strongly correlated bath.
Finally, the two-impurity sector can directly address bipolaron formation and transport by mapping the binding energy and effective mass
across the $(U_{\mathrm{ib}}/t,\,U_{\mathrm{b}}/t)$ plane; an additional direct impurity--impurity interaction would further allow one to disentangle
bare impurity-impurity interactions from bath-mediated interactions---all within the same sign-problem-free QMC framework.
Further discussion of experimental realizations and detection protocols is provided in our companion paper~\cite{chaoPRB}.

\noindent\textbf{Acknowledgments--}
We are grateful to Guido Pupillo for bringing this idea to our attention and for helpful discussions. We thank Nikolay Prokof'ev for insightful discussions.

\bibliography{impurity}
\end{document}